\documentclass[aps,prl,twocolumn,showpacs,10pt,floatfix,superscriptaddress,longbibliography]{revtex4-1}
\usepackage{lipsum}  
\usepackage[english]{babel}
\usepackage[utf8]{inputenc}
\usepackage[T1]{fontenc}

\usepackage{amsmath}
\usepackage{braket}
\usepackage{amssymb}
\usepackage[normalem]{ulem}
\usepackage{dsfont}
\usepackage{bm}
\usepackage{bbm}
\usepackage[dvipsnames]{xcolor}
\usepackage{stmaryrd}
\usepackage{mathtools}
\usepackage{physics}

\usepackage{graphicx}% Include figure files
\usepackage{dcolumn}% Align table columns on decimal point
\usepackage{hyperref}
\hypersetup{
    colorlinks=true,
    citecolor=Green,
    linkcolor=Maroon,
    urlcolor=Blue
    }

\usepackage[capitalize]{cleveref} % clever equation references
\usepackage{tikz}
\usetikzlibrary{quantikz2}

\makeatletter
\newcommand{\thickhline}{%
\noalign {\ifnum 0=`}\fi \hrule height 1.2pt
\futurelet \reserved@a \@xhline%
}
\newcolumntype{"}{@{%
\hskip\tabcolsep\vrule width 1.2pt\hskip\tabcolsep}%
}
\makeatother

\newcommand{\HH}{{H}}
\newcommand{\hh}{{h}}
\newcommand{\UU}{{U}}

\newcommand{\sig}{{\sigma}}

\begin{document}

%============================================================================

\preprint{APS/123-QED}

\title{Observing quantum many-body dynamics in emergent curved spacetime using programmable quantum processors}

\author{Brendan Rhyno}
% \email{rhyno@iqo.uni-hannover.de}
\thanks{These authors contributed equally to this work.}
\affiliation{Department of Physics, University of Illinois at Urbana-Champaign, Urbana, Illinois 61801, USA}
\affiliation{Institute of Quantum Optics, Leibniz Universität Hannover, Welfengarten 1, 30167 Hannover, Germany}

\author{Bastien Lapierre}
\thanks{These authors contributed equally to this work.}
\affiliation{Department of Physics, Princeton University, Princeton, New Jersey, 08544, USA}
\affiliation{Philippe Meyer Institute, Physics Department, École Normale Supérieure (ENS), Université PSL, 24 rue Lhomond, F-75231 Paris, France}

\author{Smitha Vishveshwara}
\affiliation{Department of Physics, University of Illinois at Urbana-Champaign, Urbana, Illinois 61801, USA}

\author{Khadijeh Najafi}
\affiliation{IBM Quantum, IBM T.J. Watson Research Center, Yorktown Heights, 10598, USA}
\affiliation{MIT-IBM Watson AI Lab, Cambridge MA, 02142, USA}

\author{Ramasubramanian Chitra}
\affiliation{Institute for Theoretical Physics, ETH Zurich, Wolfgang-Pauli-Str. 27, CH-8093 Zurich, Switzerland}

\date{\today}

%============================================================================

\begin{abstract}
We digitally simulate quantum many-body dynamics in emergent curved backgrounds using 80 superconducting qubits on IBM Heron processors.
By engineering spatially varying couplings in the spin-$\frac12$ XXZ chain, consistent with the low-energy description of the model in terms of an inhomogeneous Tomonaga-Luttinger liquid, we realize excitations that follow geodesics of an effective metric inherited from the underlying spatial deformation.
Following quenches from Néel and few-spin-flip states, we observe curved light-cone propagation, horizon-induced freezing in the local magnetization, and position-dependent oscillation frequencies set by the engineered spatial deformation. Despite strong spatial inhomogeneity, unequal-time correlators reveal ballistic quasiparticle propagation in the spin chain. These results establish large-scale digital quantum processors as a flexible platform for detailed and controlled exploration of many-body dynamics in tunable and synthetic curved spacetimes.
\end{abstract}

%============================================================================

%\keywords{Suggested keywords}%Use showkeys class option if keyword
                              %display desired

%============================================================================
\maketitle
%============================================================================

\noindent\textbf{Introduction.}
The present-day description of gravitation and cosmology hinges on curved spacetime. Against this backdrop, the full-fledged physics of black holes and the early universe relies on the coupling between geometry and quantum fields. W. G. Unruh’s seminal work~\cite{Unruh1981} unveiled the possibility of simulating field theories in curved spacetime using tabletop experiments, thereby paving the way for controlled studies of the interplay between spacetime curvature and quantum matter. These insights have spurred experimental realizations of curved spacetime analogs in both classical and quantum fluids in a range of systems~\cite{Lahav2010, Weinfurtner2011, Steinhauer2014, Eckel2018, Hu2019, Munoz_de_Nova2019, Wittemer2019, Torres2020, Banik2022, Jacquet2022, Viermann2022, Falque2025}. Here we present large-scale digital quantum processors as a fertile arena for realizing quantum dynamics in emergent curved backgrounds.

\begin{figure}[t!]
    \centering
    \includegraphics[width = \columnwidth]{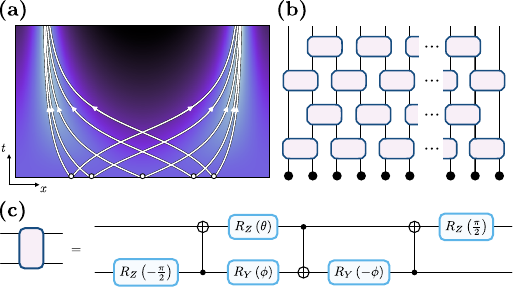}
    \caption{
    \textbf{Simulating spatially inhomogeneous dynamics with quantum processors.}
    (a) In our inhomogeneous quench protocol, the initial state acts as a source of pairs of quasiparticles (illustrated as white dots), which propagate along geodesics on a curved background.
    (b) Quantum circuit implementation of the time evolution operator $\exp(-i \delta t \HH)$ for the deformed XXZ spin chain, \cref{eq:deformedXXZ}, using a first-order Suzuki-Trotter decomposition with a time step of duration $\delta t$.
    Each Trotter layer consists of an ``odd'' and ``even'' sublayer composed of unitaries that couple nearest-neighbour qubits.
    (c) The corresponding two qubit quantum circuit; here the rotation gate angles are $\theta = 2 \delta t v_j \Delta + \pi / 2$ and $\phi = - 2 \delta t v_j - \pi / 2$ encoding both local interactions through the anisotropy $\Delta$ and emergent spacetime curvature through the deformation profile $v_j$.
    }
    \label{fig:intro_illustration}
\end{figure}

\begin{figure*}[t!]
    \centering
    \includegraphics[width = \textwidth]{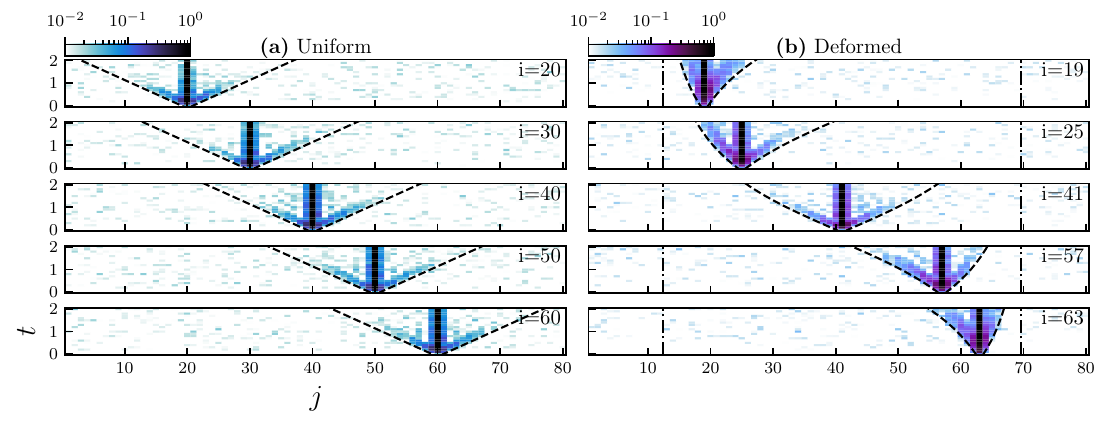}
    \caption{
    \textbf{Curved light-cone propagation.} Light-cone propagation in the two-point correlation function $|C_{ij}^{zz}(t)|$ following a quench with \cref{eq:deformedXXZ} from an initial Néel state in an interacting chain of $N=80$ qubits. (a) Uniform XXZ chain simulated on \texttt{ibm\_fez} with $\Delta=1/2$. (b) Deformed XXZ chain simulated on \texttt{ibm\_marrakesh} with the same anisotropy ($\Delta=1/2$) and a deformation given by \cref{eq:deformationprofileXXZ},
    with Rindler horizons located at $j_*+1$ and $N+1-j_*$ with
    $j_*=N/7$ (indicated with black dash-dot vertical lines).
    The observed light cones are compared to the geodesics of the metric, \cref{eq:curvedspacetimemetric} (black dashed curves) for various initial positions $i$,
    computed using $ \frac{1}{8} \int_{i}^{j} \text{d}x \, v(x)^{-1}$.
    Color maps share a common logarithmic scale, with the lower bound set by the order of magnitude associated with the standard error in each observable.
    }
    \label{fig:curvedlight-cone_propagation}
\end{figure*}

Our key guiding principle is as follows: A central feature of spacetime is its causal structure, manifested in light-cones that set bounds on the propagation of physical information. In relativistic theories, such a bound is fundamentally set by the speed of light. 
In nonrelativistic quantum many-body lattice systems, the propagation of correlations is limited by the Lieb-Robinson bound~\cite{Lieb1972}.
Quite generally, a broad class of quantum systems at criticality exhibit an emergent Lorentz invariance at large distances, so that their low-energy dynamics is governed by relativistic quantum field theories in flat spacetime with effective speeds of light.
In one spatial dimension, these theories are universal and conform to the paradigm of Tomonaga-Luttinger liquids (TLLs)~\cite{Haldane1981}.
The predicted light-cone-like propagation of quantum information~\cite{Calabrese2005,Calabrese2006} has been observed in ultracold atom experiments~\cite{Cheneau2012,Langen2013,Jurcevic2014,Tajik2023}. 
In recent years, superconducting quantum processors have also emerged as a promising complementary route for simulating the physics of quantum fields in curved spacetime~\cite{Shi2023,Sabin2023,Maceda2025,Jiang2025}.
Digital quantum simulations of large many-body systems offer advantages due to their high degree of tunability~\cite{Fauseweh2024,Smith2019,Mi2022,Farrell2024,Miessen2024,Hayata2025,Rhyno2025,Koyluouglu2026}, which enables near arbitrary control of the spatial and temporal structure of the effective curved background metric.

In this work, we simulate quantum many-body dynamics in a curved background, leveraging the superconducting transmon-qubit architecture of IBM Heron processors. Unlike analog quantum simulators, this digital platform provides fine-grained control over the effective spacetime metric, enabling the programmable realization of curved geometries with emergent horizons, and allowing a systematic exploration of their consequences for quantum dynamics.
As a concrete demonstration, we study the nonequilibrium evolution of a spatially deformed one-dimensional interacting quantum spin chain, where the deformation provides a lever to tune the local spacetime metric. The high tunability of quantum gates allows us to access a wide range of spatial deformations and interaction strengths, spanning both gapless and gapped phases. Using chains of 80 qubits, we observe deformed light-cone propagation in the gapless regime, consistent with the long-wavelength description of the spin chain as a TLL in a curved spacetime with metric
\begin{equation}
\label{eq:curvedspacetimemetric}
\text{d}s^2= \text{d}x^2-v(x)^2\text{d}t^2.
\end{equation}
Strikingly, these results are obtained with only minimal error mitigation, and the relevant dynamical signatures remain clearly visible for up to 20 Trotter steps, highlighting the robustness of digital quantum processors in simulating curved geometries.

\noindent\textbf{Setup.} 
To simulate quantum many-body dynamics in a curved background, we consider a spatially deformed version of the paradigmatic spin-$\frac12$ XXZ chain [we set $J=\hbar =1$ thereafter],
\begin{equation}
\label{eq:deformedXXZ}
H = J\sum_{j=1}^{N-1}v_j\left[\sigma_{j}^x\sigma_{j+1}^x+\sigma_{j}^y\sigma_{j+1}^y+\Delta\sigma_{j}^z\sigma_{j+1}^z\right],
\end{equation}
where the deformation profile $v_j$ is assumed to vary smoothly over mesoscopic length scales much larger than the lattice spacing. As in the homogeneous case, for smooth deformations, the model is gapless for $|\Delta|<1$ and gapped otherwise~\cite{Dubail2017_2}.
Using bosonization, it can be seen that the gapless regime lends itself to a low-energy effective description of \cref{eq:deformedXXZ} in terms of an \textit{inhomogeneous} TLL~\cite{Dubail2017_1,Dubail2017_2,Gawedzki2018,Dubail2020,Moosavi2021,Tajik2023}
\begin{equation}
H_{\text{TLL}}[v]= \int_0^L \text{d}x \frac{v(x)}{2\pi} \left[\frac{1}{K}(\partial_x \varphi(x))^2+K(\partial_x \theta(x))^2\right],
\end{equation}
for the dual bosonic fields $\varphi(x)$ and $\theta(x)$, and Luttinger parameter $K$ \footnote{We note that in the most general case, inhomogeneous TLLs can have both spatially dependent velocity $v(x)$ and Luttinger parameter $K(x)$. In order for the model to retain conformal invariance, we choose to have only $v(x)$ to be inhomogeneous, which corresponds to having a spatially uniform anisotropy $\Delta$ in the XXZ model.}.
The factor $v(x)$ can be absorbed in the spacetime metric through a conformal transformation, leading to a theory in curved spacetime with the metric in \cref{eq:curvedspacetimemetric}~\cite{Dubail2017_1,Dubail2017_2,Bermond2024}.
Therefore, low-energy gapless excitations in the spin chain, \cref{eq:deformedXXZ} with $|\Delta|<1$, which consist of left- and right-moving modes, follow lightlike geodesics, as given by $\text{d}s^2=0$.
In the following, we consider the deformation profile 
\begin{equation}
\label{eq:deformationprofileXXZ}
v_j = \frac{\sin(\frac{\pi}{N}(j-1-j_*))\sin(\frac{\pi }{N}(j-1+j_*))}{\sin(\frac{2\pi }{N}j_*)}.
\end{equation}
With this choice of deformation, the resulting spacetime metric, \cref{eq:curvedspacetimemetric}, features two emergent Rindler horizons around sites $j_*+1$ and $N+1-j_*$, where the velocity approaches zero linearly.
Consequently, between the two horizons, every chiral (antichiral) gapless quasiparticle accumulates at the site nearest to $j_* + 1$ ($N+1-j_*$),
which leads to a localization of all quantum entanglement at the two horizons~\cite{Fan2020}. 
The resulting deformed Hamiltonian admits a natural interpretation as the entanglement Hamiltonian in (1+1)d conformal field theory~\cite{Cardy2016, Zhu2020},
and can also be realized as an effective Hamiltonian via Floquet engineering~\cite{Lapierre2020, Fan2020, Lapierre2021}. Note that all quasiparticles (and hence energy) accumulate at the two horizons with time. This can be interpreted as an effective cooling of the bulk and can be harnessed as a route towards efficient ground state preparation~\cite{Kuzmin2022, Wen2022}.

\begin{figure}
    \centering
    \includegraphics[width = \columnwidth]{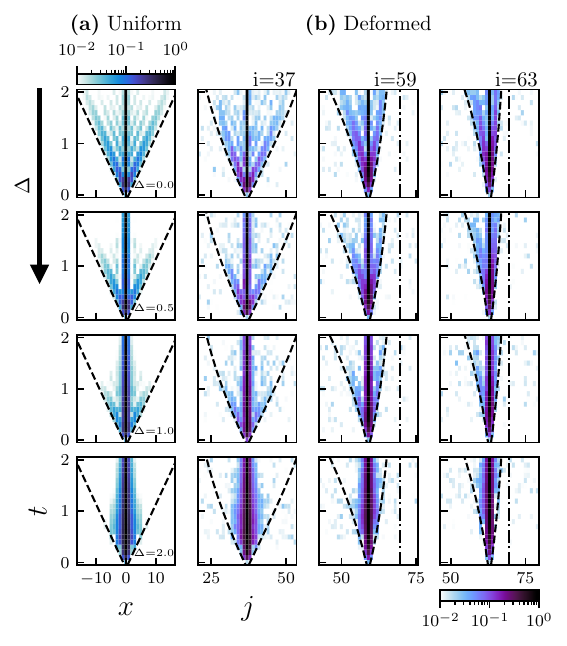}
    \caption{
    \textbf{Tuning interactions.} (a) Spatially averaged two-point correlator
    $|C^{zz}(x,t)| = |\frac{1}{N_x}\sum_i C_{i,i+x}^{zz}(t)|$, where $N_x$ is the number of pairs of sites separated by distance $|x|$~\cite{Keesling2019},
    following a quench with the uniform XXZ chain initialized in a Néel state using several interaction strengths, $\Delta = 0, 1/2, 1,$ and $2$.
    (b) Correlation function $|C_{ij}^{zz}(t)|$ following a quench with the inhomogeneous XXZ chain (same deformation as in \cref{fig:curvedlight-cone_propagation}(b)) for different positions, 
    $i=37,59,63$, using the same initial state and $\Delta$ values as in (a).
    All quantum simulations here were performed on \texttt{ibm\_fez}.
    }
    \label{fig:different_Delta}
\end{figure}

Our nonequilibrium protocol is an inhomogeneous variant of the paradigmatic global quantum quench~\cite{Calabrese2016}: we initialize the system in a massive state $|\Psi_0\rangle$, and extract the time evolution of observables after quenching with the gapless deformed XXZ model, \cref{eq:deformedXXZ}. At $t=0$, pairs of excitations with opposite chirality are created at all positions and propagate according to geodesics of the metric, \cref{eq:curvedspacetimemetric}.
In the following, (i) we experimentally extract signatures of the curved metric, \cref{eq:curvedspacetimemetric}, and (ii) we demonstrate the ballistic nature of the excitations of \cref{eq:deformedXXZ}, confirming that the dynamics of such an inhomogeneous interacting spin chain is governed by gapless quasiparticles.

\noindent\textbf{Curved light-cone propagation.} As a direct probe of the effective curved metric generated by the spatially dependent spin coupling in \cref{eq:deformedXXZ}, we simulate the equal-time connected spin-spin correlation function, defined as
\begin{equation}
\label{eq:correlator_zz}
C^{zz}_{ij}(t)= \langle \sigma^z_i(t)\sigma^z_j(t) \rangle -\langle \sigma^z_i(t)\rangle\langle\sigma^z_j(t) \rangle,
\end{equation}
starting from an initial Néel state
$|\Psi_0 \rangle = |\uparrow\downarrow\uparrow\downarrow\dots \rangle$.
This choice has two advantages: it circumvents the challenging problem of preparing the critical ground state, and, more importantly, as the time evolution is fundamentally dictated by the low-energy spinons stemming from domain wall formation, it directly probes the effective Luttinger theory.
Working in the Schr\"odinger picture, the initial state time evolves as $\ket{\Psi(t)}=e^{-i t H} \ket{\Psi_0}$, where the time evolution operator is achieved by a first-order Suzuki-Trotter decomposition using time steps of duration $\delta t = 0.1$ (see Supplementary Material (SM)~\cite{supmat} for details and references~\cite{Vatan2004,Smith2019,Keenan2023}).
After initial state preparation and applying gate implementations of the Trotterized unitary to reach the $s=0,1,2,\dots,20$ Trotter step, the system is measured along the $z$-axis (computational basis), and this process is repeated many times.
Here, we use $2^{14}$ measurement shots, which results in a standard error in the correlation functions on the order of $10^{-2}$.
These quantum simulations were carried out on the IBM Heron processors \texttt{ibm\_fez} and \texttt{ibm\_marrakesh}, using the Qiskit Runtime Sampler primitive. Circuit layouts were compiled with level-3 transpiler optimization, and Pauli twirling together with dynamical decoupling were applied as low-cost error mitigation and suppression techniques.

\begin{figure*}[!htbp]
    \centering
    \includegraphics[width = 0.983 \textwidth]{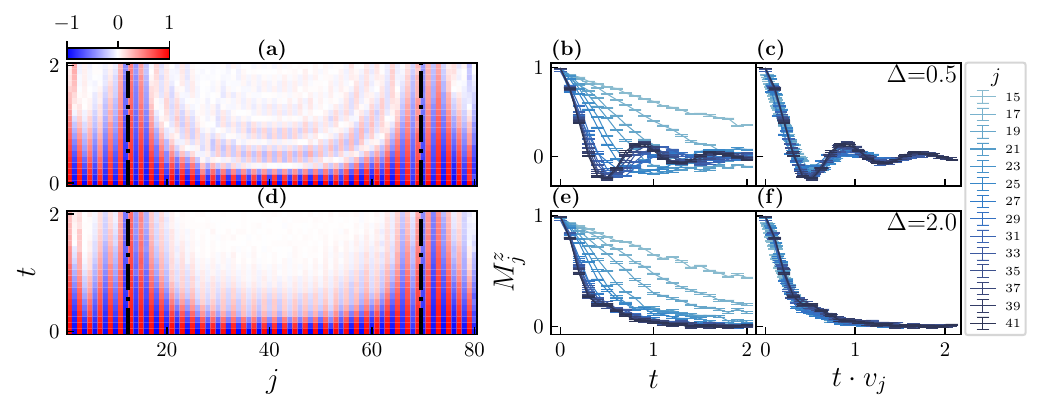}
    \caption{
    \textbf{Magnetization dynamics.} (a) Time evolution of the local magnetization $M^z_j(t)$ from an initial Néel state, for the same parameters as \cref{fig:different_Delta}(b). The effective velocity profile in the metric, \cref{eq:curvedspacetimemetric}, manifests as inhomogeneous oscillation periods that freeze at the horizons (denoted by black dash-dot vertical lines). (b) Damped oscillations at different initial positions reveal a strong position dependence of the frequency. (c) Collapse of the magnetization curves after rescaling the time axis by the spatial deformation.
    Standard errors in each measurement have been included.
    (d-f) Corresponding results in the gapped phase ($\Delta=2$), where oscillations are suppressed.
    }
    \label{fig:magnetization_dynamics}
\end{figure*}

The results for the correlator in \cref{eq:correlator_zz} are displayed in \cref{fig:curvedlight-cone_propagation}(a) for a uniform XXZ chain, and \cref{fig:curvedlight-cone_propagation}(b) for the inhomogeneous chain; the latter are explicitly compared with the lightlike geodesics of the metric, \cref{eq:curvedspacetimemetric}, showing excellent agreement.
Our results confirm that the light-cone structure is strongly sensitive to the location of the initial excitation in the deformed setting. Notably, a clear left-right asymmetry emerges as the excitations approach the Rindler horizons generated by the deformation in \cref{eq:deformationprofileXXZ}.
We emphasize that, although decoherence in quantum processors prevents us from accessing times at which gapless quasiparticles traverse the full system, simulating sufficiently large system sizes $N$ is nonetheless essential. Only then do the couplings in \cref{eq:deformedXXZ} vary slowly enough relative to the lattice spacing for the low-energy dynamics to admit an emergent field-theoretic description in curved spacetime.

To probe how interactions affect the curved light-cone structure, we simulated the inhomogeneous XXZ chain across different anisotropy values $\Delta$ spanning both critical and gapped regimes; the corresponding results are displayed in \cref{fig:different_Delta}. Clear resonances manifest within the light cones, consistent with coherent free propagation in the noninteracting case ($\Delta=0$). The light cones remain stable as we tune $\Delta$ through the gapless regime. In the gapped phase ($\Delta=2$), we observe a local deformation-dependent exponential suppression of propagation.

\noindent\textbf{Magnetization as a probe of local curvature.} 
For the uniform XXZ model in the gapless regime, it has been demonstrated that the dynamics of the magnetization starting from a Néel state shows damped Rabi-like oscillations, with frequency set by the exchange coupling $J$. Importantly, these oscillations lie beyond the TLL description~\cite{Barmettler2010, Jepsen2020}, reflecting their non-universal, lattice-scale origin. It is therefore natural to ask whether such oscillations are affected by the effective metric, \cref{eq:curvedspacetimemetric}.
We demonstrate that this is the case by studying the time dependence of the local magnetization 
$M^z_j(t)=\langle \sigma^z_j(t) \rangle$ 
after an inhomogeneous quench with \cref{eq:deformedXXZ}, starting from the Néel state.
While the absence of translational invariance would in general preclude any nontrivial structure of the magnetization beyond exponential decay, the smooth inhomogeneity in \cref{eq:deformedXXZ} leads to intriguing dynamics.
In fact, as we show in the SM~\cite{supmat} for the noninteracting case $\Delta=0$, $M^z_j(t)$ showcases robust oscillations, similar to the homogeneous case, but with an explicitly spatially dependent frequency $\omega_j \propto J|v_j|$.
A direct consequence of this relation is that the local magnetization freezes near the horizons, accompanied by a divergence in the decay time. In other words, the memory of the initial state is kept for parametrically long times when approaching the horizons.

We confirm the emergence of this spatially dependent dynamics in the magnetization $M_j^z(t)$ for $|\Delta| < 1$, as shown in \cref{fig:magnetization_dynamics}(a--c). Focusing on $\Delta = 1/2$, we measure $M_j^z(t)$ for an initial Néel state and observe the expected damped oscillations, whose frequency varies with position in the gapless regime. As demonstrated in \cref{fig:magnetization_dynamics}(c), these curves collapse onto a single universal profile once the time axis is rescaled by the spatial deformation $v_j$. 
This implies that oscillations of the magnetization directly probe the local velocity, \cref{eq:curvedspacetimemetric}.
Remarkably, our results indicate that the effective metric governs the observable dynamics even beyond the continuum regime.

In contrast, in the gapped regime with $\Delta = 2$, we observe that the local magnetization does not exhibit oscillations, as shown in \cref{fig:magnetization_dynamics}(d--e). Nevertheless, even for $\Delta = 2$, the magnetization remains spatially modulated by the deformation, and the decay rate is parametrically reduced near the horizons.
Remarkably, the magnetization decay curves show an almost perfect collapse once time is rescaled by the spatial deformation, as shown in \cref{fig:magnetization_dynamics}(f).

\begin{figure}[!htbp]
    \centering
    \includegraphics[width = 0.964 \columnwidth]{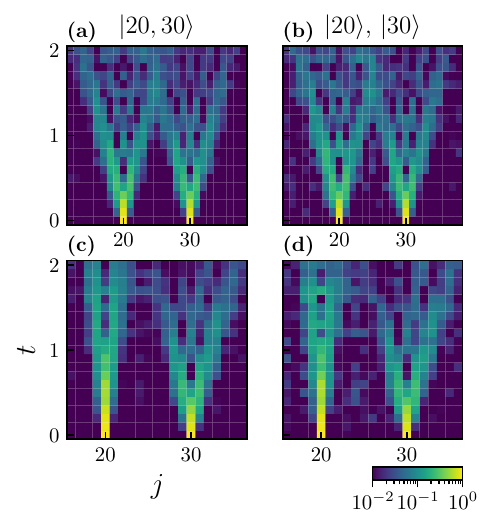}
    \caption{\textbf{Ballistic spreading of correlations.}
    Panels (a) and (c) show $N=80$ qubit simulation results of the unequal-time correlation function for a uniform chain and a deformed chain ($\Delta=1/2$), respectively, when the initial state contains two spin flips located at sites 20 and 30.
    Panels (b) and (d) instead compare the dynamics starting from a single spin flip: we prepare one state with the flip at site 20 and another at site 30, compute the corresponding correlation functions, and overlay the results for direct comparison with the two-spin flip state.
    In all cases we have subtracted off the background magnetization by performing simulations with an initial all-up spin state.}
    \label{fig:ballisticcorrelator}
\end{figure}

\noindent\textbf{Signatures of ballistic transport.} In the absence of inhomogeneity, the XXZ model is integrable and exhibits ballistic transport for $|\Delta| < 1$~\cite{Giamarchi2003}. Although generic deformations of the XXZ model break integrability and typically lead to diffusive transport~\cite{Prosen2009}, the spin chain, \cref{eq:deformedXXZ}, remains ballistic and is described in the low-energy regime by a TLL on a curved background.
We now test the ballistic nature of our model by studying the propagation of quasiparticles from the unequal-time correlator $G^{zz}_{ji}(t,0) = \langle \sigma^z_j(t)\sigma^z_i(0) \rangle$.
The initial states we consider in the following are of the form
\begin{align}
\label{eq:spinflip2}
    | j_1,j_2,\dots,j_n  \rangle
    &= \prod_{a=1}^{n} \sig^x_{j_a}  | \uparrow\dots\uparrow \rangle
    ,
\end{align}
where the spins at the sites $j_1,j_2,\dots,j_n$ are down, with the rest being up.
We consider two distinct protocols: one where the initial state consists of a double spin-flip, i.e. $| j_1,j_2 \rangle$, and another with single spin-flip states of the form $| j_1 \rangle$ and $| j_2 \rangle$.
For these initial states, the unequal-time spin-spin correlator is related to the local magnetization by
$G^{zz}_{ji}(t,0) = M^z_j(t) M^z_i(0)$.
In the case of a uniform XXZ chain, we find that the unequal-time correlator is nearly identical between both initial state choices, as shown in \cref{fig:ballisticcorrelator}(a-b).
This independence of the initial state is a clear signature of ballistic transport in the gapless chain. For the deformed XXZ chain, we similarly observe independence from the initial spin flips; see \cref{fig:ballisticcorrelator}(c-d).
Due to the strong effects of the deformation, the quasiparticles emitted at positions $20$ and $30$ do not significantly overlap on the simulable timescale.

\noindent\textbf{Outlook.} Our successful demonstration of critical quantum dynamics in a curved background using digital simulators heralds exciting future prospects.
By changing the connectivity of our qubit graph, our work can be extended to two dimensions in a straightforward manner.
The ease of imprinting time-dependent metrics in these platforms is of direct relevance to quantum simulations of cosmological phenomena such as inflationary particle production and density fluctuations~\cite{Fedichev2004,Jain2007,Prain2010,Schmiedmayer2013,Steinhauer2022,TolosaSimeon2022,Bhardwaj2024}, analog black holes~\cite{Unruh1981,Garay2000,Novello2002,Visser2002,Steinhauer2016}, and other realms of analog gravity~\cite{Barcelo2011,Volovik2009}.
Beyond the direct simulation of quantum field theories, these simulators offer a platform to investigate fundamental questions, such as the emergence of inhomogeneous butterfly velocity in out-of-time-order correlators~\cite{Swingle2018,Swingle2016,Dag2019,Lapierre2025,Hayata2025} and Kardar–Parisi–Zhang (KPZ) scaling from noisy spatiotemporal deformations~\cite{Bernard2020,Keenan2023}.
A promising future direction is the coupling of the original metric to the quantum evolution of matter via backreaction feedback to update inhomogeneities of the system.
This permits the simulation of toy models of gravity~\cite{Oppenheim2023} and to test if horizons emerge dynamically from backreaction~\cite{Almheiri2015,Schwartz2019}.

\begin{acknowledgments}\noindent\textbf{Acknowledgments.} 
B.L. acknowledges financial support from the Swiss National Science Foundation (Postdoc.Mobility Grant No. 214461).
%and from the Philippe Meyer Institute at Ecole Normale Supérieure.
B.R. acknowledges the support of the Quantum Leap Challenge Institute for Hybrid Quantum Architectures and Networks Grant No. OMA2016136.
S.V. acknowledges the support of the National Science Foundation through Grant No. DMR2004825.
We acknowledge the use of IBM Quantum Credits for this work. The views expressed are those of the authors, and do not reflect the official policy or position of IBM or the IBM Quantum team.
\end{acknowledgments}

\noindent\textbf{Competing Interests.} The authors declare no competing interests.

\bibliography{ref}% Produces the bibliography via BibTeX.

\onecolumngrid 
\clearpage
\makeatletter 

\begin{center}
	\textbf{\large Supplementary Material for ``Observing quantum many-body dynamics in emergent curved spacetime using programmable quantum processors''}\\
	[1em]
	Brendan Rhyno$^{1,2}$, Bastien Lapierre$^{3,4}$, Smitha Vishveshwara$^{1}$, Khadijeh Najafi$^{5,6}$, and Ramasubramanian Chitra$^{7}$ \\[.1cm]
	{\itshape \small ${}^1$Department of Physics, University of Illinois at Urbana-Champaign, Urbana, Illinois 61801, USA \\ 
	${}^2$Institute of Quantum Optics, Leibniz Universität Hannover, Welfengarten 1, 30167 Hannover, Germany\\
    ${}^3$Department of Physics, Princeton University, Princeton, New Jersey, 08544, USA\\
    ${}^4$Philippe Meyer Institute, Physics Department, École Normale Supérieure (ENS), Université PSL, 24 rue Lhomond, F-75231 Paris, France\\
    ${}^5$ IBM Quantum, IBM T.J. Watson Research Center, Yorktown Heights, 10598, USA\\
    ${}^6$ MIT-IBM Watson AI Lab, Cambridge MA, 02142, USA\\
    ${}^7$Institute for Theoretical Physics, ETH Zurich, Wolfgang-Pauli-Str. 27, CH-8093 Zurich, Switzerland}\\
	(Dated: \today)\\[1cm]
\thispagestyle{titlepage} 
\end{center} 	

% --- Reset counters for S-numbering ---
%\renewcommand{\theequation}{S\arabic{equation}}
\renewcommand{\thefigure}{S\arabic{figure}}
\renewcommand{\theHfigure}{S\arabic{figure}}
\setcounter{figure}{0}

% --- Start appendices so sections are A, B, C ---
\appendix
\setcounter{secnumdepth}{2} % ensure sections are numbered/lettered

% --- Add standalone title without killing numbering ---
%\section*{Supplementary Material}
\addcontentsline{toc}{section}{Supplementary Material}
This Supplemental Material comprises several appendices containing technical details and additional data that support the results presented in the main text.
Appendix~\ref{app:simulating_spin_chains} details the implementation of the inhomogeneous XXZ spin chain on IBM quantum processors.
Appendix~\ref{app:error_analysis} details the error analysis performed on measured quantities.
Appendix~\ref{app:localmagnetization} presents further data on the local magnetization, including results obtained in the absence of deformation or in the noninteracting limit.
Appendix~\ref{app:ballsiticspread} provides additional data on the ballistic spreading of excitations. Finally, Appendix~\ref{app:twopoint} reports comprehensive quantum-simulation results for the two-point correlators.

\section{Simulating spin chain dynamics using IBM quantum computers}
\label{app:simulating_spin_chains}

In this appendix, we provide details on how initial states are prepared, time evolution with the XXZ model is performed, and correlators are extracted using IBM Quantum hardware.

\textit{Initial state preparation}:
By default, the $N$-qubit quantum circuits that we study with IBM Quantum hardware are initialized in the all-up state $
\ket{\uparrow\dots\uparrow} = \ket{\uparrow}\otimes\cdots\otimes\ket{\uparrow}$.
In this work, we consider initial product states of the form
\begin{align}
    \label{eq:initial_states}
    | j_1,j_2,\dots \rangle
    &= | \uparrow\dots\underset{j_1}{\downarrow}\dots\underset{j_2}{\downarrow}\dots\uparrow \rangle
    ,
\end{align}
where the spins at the sites $j_1,j_2,\dots$ are down, with the rest being up.
We will be particularly interested in the Néel state,
$|\uparrow\downarrow\uparrow\downarrow\dots\uparrow\downarrow \rangle$, which consists of alternating up and down spins.
The initial-state preparation consists of a single gate layer to flip each spin at the desired sites by applying the Pauli-X gate,
\begin{align}
\ket{\downarrow} =
\begin{quantikz}
\lstick{$|{\uparrow}\rangle$} & \gate{X} &
\end{quantikz}
.
\end{align}

\textit{Time evolution}: 
For a static Hamiltonian, unitary time evolution is achieved by acting the operator $\UU(t) = e^{-i t \HH} $ on the initial state $\ket{\Psi_0}$.
Here, we simulate dynamics in discrete steps of constant duration $\delta t$.
Namely, to evolve the system to time $t_s = s \delta t$ with $s$ a nonnegative integer, we apply the operator
\begin{align}
    \label{eq:discrete_time_evolution}
    \UU(s \delta t)
    = \underbrace{
    e^{-i \delta t \HH} \dots e^{-i \delta t \HH} }_\text{s times}
    .
\end{align}

To simulate time evolution in the inhomogeneous XXZ model, \cref{eq:deformedXXZ}, we utilize the optimal decomposition with regard to two qubit gates in~\cite{Vatan2004,Smith2019,Keenan2023}:
\begin{align}
\label{eq:XYZ_unitary}
{N}_{j_1,j_2}(\alpha,\beta,\gamma)
&\equiv e^{ i (\alpha \sig^{x}_{j_1} \sig^{x}_{j_2} + \beta \sig^{y}_{j_1} \sig^{y}_{j_2} + \gamma \sig^{z}_{j_1} \sig^{z}_{j_2} ) }
\nonumber\\
&= e^{i \pi/4} \cdot
\begin{quantikz}
\lstick{$j_1$} & \qw & \targ{} & \gate{R_Z\left(\frac{\pi}{2}-2\gamma\right)} & \ctrl{1} & \qw & \targ{} & \gate{R_Z\left(\frac{\pi}{2}\right)} &
\\
\lstick{$j_2$} & \gate{R_Z\left(-\frac{\pi}{2}\right)} & \ctrl{-1} & \gate{R_Y\left(2\alpha - \frac{\pi}{2}\right)} & \targ{} & \gate{R_Y\left(\frac{\pi}{2}-2\beta\right)} & \ctrl{-1} & \qw &
\end{quantikz}
,
\end{align}
where the top (bottom) wire refers to qubit $j_1$ ($j_2$). The single qubit gates ``$R_\alpha(\theta)$'' rotate a qubit about the $\alpha$-axis by angle $\theta$. The vertical connected wires represent controlled-X gates or ``CNOT'' gates where the solid dot refers to the control qubit and the circled plus refers to the target qubit~\cite{Nielsen2001}.
There is a global phase factor that appears on all qubits and can be safely disregarded~\cite{Vatan2004}.
Also note that using the top wire for qubit $j_1$ and the bottom wire for qubit $j_2$ was an arbitrary choice as the circuit is invariant under swapping the two qubits~\cite{Smith2019}.

For each time step, we approximate the operator $e^{-i \delta t \HH}$ for the Hamiltonian in \cref{eq:deformedXXZ} using a first-order Suzuki-Trotter decomposition~\cite{Trotter1959,Suzuki1993}:
\begin{align}
    \label{eq:Trotter}
    e^{-i \delta t ({A} + {B})} = e^{-i \delta t {A}} e^{-i \delta t {B}} + \mathcal{O}(\delta t^2)
    .
\end{align}
A first-order decomposition is done as higher-order Trotter breakups introduce increased gate depth~\cite{Smith2019}.
One can write the inhomogeneous XXZ Hamiltonian as $\HH = \sum_{j=1}^{N-1} \hh_j$ where
$
\hh_j \equiv v_j\left[\sig_{j}^x\sig_{j+1}^x+\sig_{j}^y\sig_{j+1}^y+\Delta\sig_{j}^z\sig_{j+1}^z\right]
$
is a local energy density.
Importantly, these local operators do not commute at adjacent sites:
\begin{align}
    [ \hh_{j_1} , \hh_{j_2} ] \ne 0 \text{ , if } {j_2}={j_1} \pm 1
    .
\end{align}
Assuming $\delta t$ is sufficiently small, we use the Suzuki-Trotter decomposition in \cref{eq:Trotter} by setting ${A} \equiv \sum_{j\in\text{even}} \hh_j$ and ${B} \equiv \sum_{j\in\text{odd}} \hh_j$~\cite{Keenan2023}.
These operators do not commute, $[{A},{B}] \neq 0$, but each local operator within ${A}$ commutes (similarly for ${B}$).
This allows us to write the Trotterized operator using products of the unitary in \cref{eq:XYZ_unitary}.
In particular, a single Trotterized time-step forward is achieved with:
\begin{align}
    \label{eq:XXZ_time_step}
    e^{-i \delta t \HH}
    &\approx
    \prod_{j\in\text{even}} {N}_{j,j+1} \left(
    - \delta t v_j,
    - \delta t v_j,
    - \delta t v_j \Delta \right)
    \prod_{j\in\text{odd}} {N}_{j,j+1} \left(
    - \delta t v_j,
    - \delta t v_j,
    - \delta t v_j \Delta \right)
    \nonumber\\
    &=
    \begin{quantikz}
    & \gate[wires=2]{{N}_{1,2}} & & \\
    & & \gate[wires=2]{{N}_{2,3}} & \\
    & \gate[wires=2]{{N}_{3,4}} & & \\
    & & \gate[wires=2]{{N}_{4,5}} & \\
    & \gate[wires=2]{{N}_{5,6}} & & \\
    & & & \\
    \end{quantikz}
    .
\end{align}
\textbf{$\Delta = 0$ case}:
To simulate a time step forward with the inhomogeneous XX model, one could use the previous result in \cref{eq:XXZ_time_step} and simply set $\Delta=0$. However, the 3 CNOT gates involved in each application of \cref{eq:XYZ_unitary} is not optimal.
Indeed, for $\Delta=0$, one can reduce the number of CNOT gates from 3 down to 2~\cite{Smith2019}.
Here, we use the circuit in~\cite{Smith2019} to implement the following XZ unitary:
\begin{align}
    \label{eq:XZ_unitary}
    {N}_{j_1,j_2}(\alpha,0,\gamma)
    =
    e^{i( \alpha \sig^{x}_{j_1} \sig^{x}_{j_2} + \gamma \sig^{z}_{j_1} \sig^{z}_{j_2} )}
    =
    \begin{quantikz}
    \lstick{$j_1$} & \targ{} & \gate{R_Z(-2\gamma)} &  \targ{} & \\
    \lstick{$j_2$} & \ctrl{-1} & \gate{R_X(-2\alpha)} & \ctrl{-1} & 
    \end{quantikz}
    .
\end{align}
Because this circuit realizes Ising XX and ZZ terms, instead of the desired XX and YY terms, we rotate all qubits 90 degrees about the $x$-axis:
\begin{align}
    \label{eq:XY_unitary}
    {N}_{j_1,j_2}(\alpha,\beta,0)
    &=
    e^{ i (\alpha \sig^{x}_{j_1} \sig^{x}_{j_2} + \beta \sig^{y}_{j_1} \sig^{y}_{j_2} ) }
    = 
    \begin{quantikz}
    \lstick{$j_1$} & \gate{R_X(\pi/2)} & \targ{} & \gate{R_Z(-2\beta)} &  \targ{} & \gate{R_X(-\pi/2)} & \\
    \lstick{$j_2$} & \gate{R_X(\pi/2)} & \ctrl{-1} & \gate{R_X(-2\alpha)} & \ctrl{-1} & \gate{R_X(-\pi/2)} &
    \end{quantikz}
    .
\end{align}
One can now proceed to construct a circuit that implements the (Trotterized) time-step forward in the exact same manner as we did for the XXZ model.
Simply use the circuit shown in \cref{eq:XXZ_time_step}, but substitute \cref{eq:XY_unitary} for the XY unitary circuit.
However, there is another simplification that can be made.
When substituting \cref{eq:XY_unitary} into \cref{eq:XXZ_time_step}, a great deal of cancellation occurs between the X-rotation gates.
Between the even and odd sub-layers of each Trotter layer, and also with the application of each additional Trotter layer,
the X-rotation gates encounter their inverse. Hence, we only apply a single layer of X-rotation gates at the start of the time evolution and then apply its inverse at the very end.

\textit{Projective measurements}:
The Qiskit Runtime Sampler primitive allows one to submit their quantum circuits through the IBM cloud to be run on backend quantum hardware.
By default, the Sampler primitive measures these circuits in the computational basis ($\ket{0}=\ket{\uparrow}$ and $\ket{1}=\ket{\downarrow}$) for a specified number of shots and returns the number of counts (i.e. the number of times a given configuration was observed).
From the probability of a configuration occurring, one can then construct the expectation values of various equal-time Pauli-Z operators (e.g. $\langle \sig^z_j(t) \rangle = \sum_{\{n_i=0,1\}} p_{\{n_i\}}(t) (1 - 2n_j)$ where $\{n_i\}$ is a computational basis configuration and $p_{\{n_i\}}(t)$ the probability of measuring the configuration at time $t$).
To extract these expectation values for multiple time-steps $s=0,1,2,\dots,P$ ($s=0$ corresponds to measurements of the initial state), one would submit the following batch of circuits to the Sampler primitive:
\begin{align}
\label{eq:time_evolution_circuit}
s=0 :&
\begin{quantikz}[wire types={b},classical gap=0.1cm]
\lstick{$\ket{\Psi_0}$} & \meter{}
\end{quantikz}
\nonumber\\
s=1 :&
\begin{quantikz}[wire types={b},classical gap=0.1cm]
\lstick{$\ket{\Psi_0}$} & \gate{e^{-i\delta t \HH}} & \meter{}
\end{quantikz}
\nonumber\\
s=2 :&
\begin{quantikz}[wire types={b},classical gap=0.1cm]
\lstick{$\ket{\Psi_0}$} & \gate{e^{-i\delta t \HH}} & \gate{e^{-i\delta t \HH}} & \meter{}
\end{quantikz}
\nonumber\\
\vdots&
\nonumber\\
s=P :&
\begin{quantikz}[wire types={b},classical gap=0.1cm]
\lstick{$\ket{\Psi_0}$} & \gate{e^{-i\delta t \HH}} & 
\push{\ \dots \ } & \gate{e^{-i\delta t \HH}} &
\push{\ \dots \ } & \gate{e^{-i\delta t \HH}} & \meter{}
\end{quantikz}
,
\end{align}
where each unitary is replaced with the previously discussed Trotter decomposition.

\section{Error analysis}
\label{app:error_analysis}

When displayed, the error bars reported on the measurements of an operator $O$ are standard errors calculated using $\delta_{\langle O \rangle} = \sqrt{\mathrm{Var}(O)/\mathrm{shots}}$ where $\mathrm{Var}(O) = \langle O^2 \rangle - \langle O \rangle^2$ is the measured variance in the operator and $\mathrm{shots}$ $(=2^{14})$ is the number of measurements used in the experiment.
For the local magnetization along the $z$-axis, $M^z_j(t) = \langle \sig^z_j(t) \rangle$, one finds
\begin{align}
    \delta_{M^z_j(t)} = \sqrt{\frac{1 - \langle \sig^z_j(t) \rangle^2}{ \mathrm{shots}}}
    .
\end{align}
For the equal-time connected spin-spin correlation function, $C^{zz}_{ij}(t) = \langle \sig^z_i(t) \sig^z_j(t) - M^z_i(t) M^z_j(t) \rangle$, one finds
\begin{align}
    \delta_{C^{zz}_{ij}(t)} = \sqrt{\frac{1 - \langle \sig^z_i(t) \sig^z_j(t) \rangle^2}{\mathrm{shots}} }
    .
\end{align}
In both cases, the maximum standard error on these local correlation functions is $\sqrt{1/\mathrm{shots}}=0.0078125$ which is of the order $10^{-2}$.

\section{Local magnetization dynamics }
\label{app:localmagnetization}

In this appendix, we provide further data on the time evolution of the magnetization for a system initialized in a Néel state, both for uniform and inhomogeneous XXZ quenches. 

\subsection{Uniform XXZ chain}

\begin{figure}[htbp]
    \centering
    \includegraphics[width = 0.5\columnwidth]{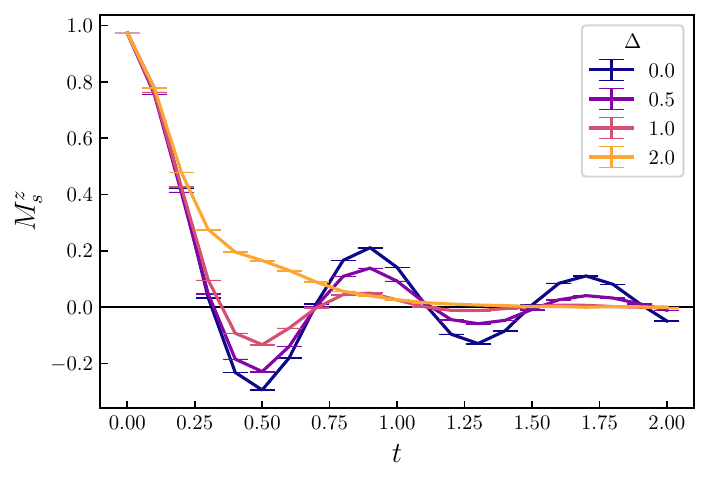}
    \caption{
    Time evolution of the staggered magnetization for a quench with the uniform XXZ chain (with $N=80$ qubits) initialized in a Néel state, with different values of the anisotropy $\Delta=0, 1/2, 1, 2$. Damped oscillations appear in the gapless phase of the XXZ model, while they are clearly absent in the gapped phase.
    The error bars are standard errors, given by
    $\delta_{M^z_s(t)} = \sqrt{\sum_{ij=n+1}^{N-n} f_i f_j \, C^{zz}_{ij}(t)/\mathrm{shots}}$ where $f_j = (-1)^{j-1} / (N-2n)$, and $C^{zz}_{ij}(t)$ is the measured equal-time connected spin-spin correlation function.
    }
    \label{fig:magnetizationuniform}
\end{figure}

Our primary focus lies on effective curved-spacetime dynamics induced by spatial deformation. For completeness, however, we also report results for magnetization dynamics in the uniform XXZ chain ($v_j \equiv 1$). In this case, as was both analytically and numerically studied in~\cite{Barmettler2010}, one expects damped oscillations of the magnetization in the gapless regime ($|\Delta|<1$) and no oscillations in the gapped regime ($|\Delta|>1$). We confirm this behavior in \cref{fig:magnetizationuniform} by studying the time evolution of the staggered magnetization, defined as
\begin{equation}
M_s^z(t) = \frac{1}{N-2n}\sum_{j=n+1}^{N-n}(-1)^{j-1} M_j^z(t),
\end{equation}
where the threshold $n=9$ was chosen to suppress the finite-size effects associated with open boundary conditions.
In particular, we find that the oscillation frequency does not directly depend on the anisotropy $\Delta$ within the gapless regime, as expected, while increasing $\Delta$ leads to a faster decay of the magnetization. These results provide an experimental verification of the tensor network calculations in Ref.~\cite{Barmettler2010}.

\subsection{Deformed XX chain (\texorpdfstring{$\Delta=0$}{Delta=0})}
\begin{figure}[htbp]
    \centering
    \includegraphics[width = 0.5\linewidth]{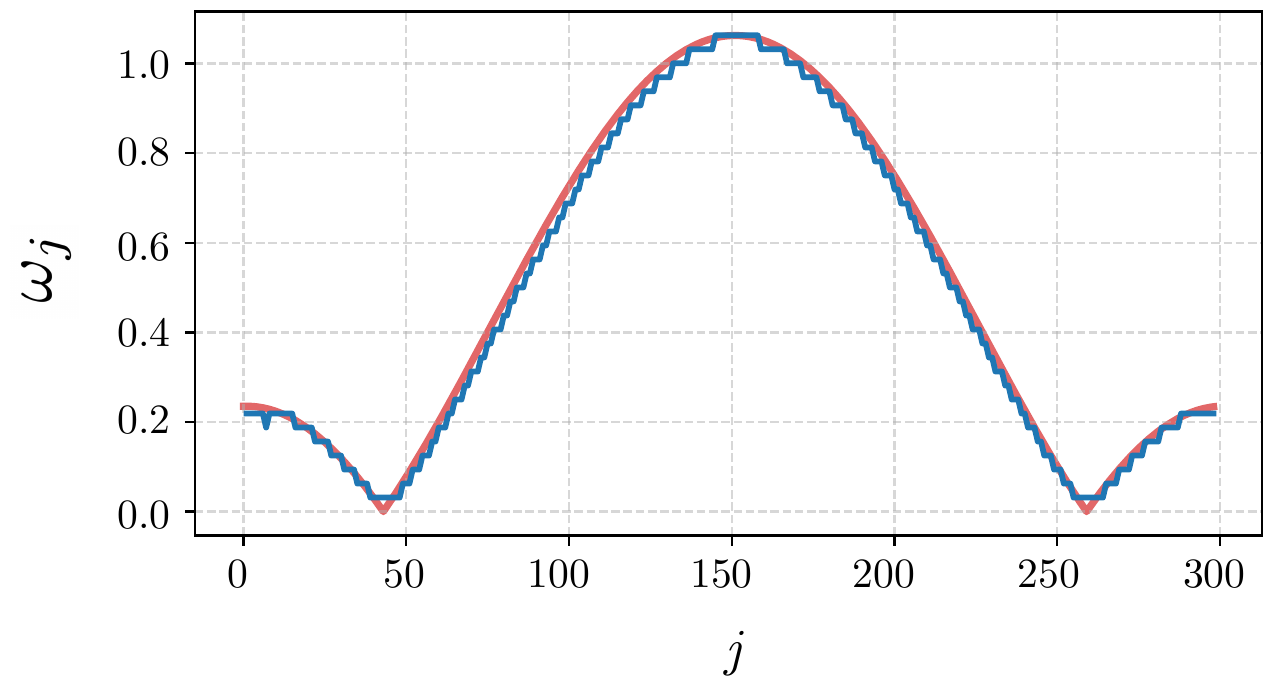}
    \caption{
    Numerical comparison between the spatially-dependent frequency $\omega_j$ in the oscillations of $\langle \sig^z_j(t) \rangle$ after a quench with \cref{eq:freefermionhamiltonian} (blue) with the absolute value deformation profile given by \cref{eq:deformationprofileXXZ} (red), for a system size of $N=300$. As is manifest from the plot, the frequency follows the local deformation profile in the thermodynamic limit.
    }
    \label{fig:lattice_profilecompa}
\end{figure}

    In this section, we provide data for the inhomogeneous noninteracting limit of the XXZ model, which we refer to as the inhomogeneous XX model. 
    This spin model can be mapped through a Jordan-Wigner transformation to a free-fermionic model of the form
\begin{equation}
\label{eq:freefermionhamiltonian}
H_{\text{free}}= 2J\sum_{j=1}^{N-1}v_j (c_{j+1}^{\dagger}c_j+h.c.).
\end{equation}
Using exact diagonalization, we compute the local magnetization $\langle \sig^z_j(t) \rangle = 1 - 2\langle n_j(t) \rangle$, starting from the product state $|0101\cdots01\rangle$. Similar to the uniform case $v_j\equiv 1$, the dynamics of $\langle \sig^z_j(t) \rangle$ displays robust oscillations in this noninteracting setting. While in the uniform case the frequency of the oscillations is set by $J$, we find that in the inhomogeneous case the frequency depends on space through $\omega_j \sim J v_j$. This behavior is illustrated in \cref{fig:lattice_profilecompa}, where the local frequency $\omega_j$ is explicitly compared to the deformation profile $v_j$. This numerical calculation shows that the local deformation profile completely determines the oscillatory dynamics of the local magnetization in the noninteracting limit.

\begin{figure}[htbp]
    \centering
    \includegraphics[width = \textwidth]{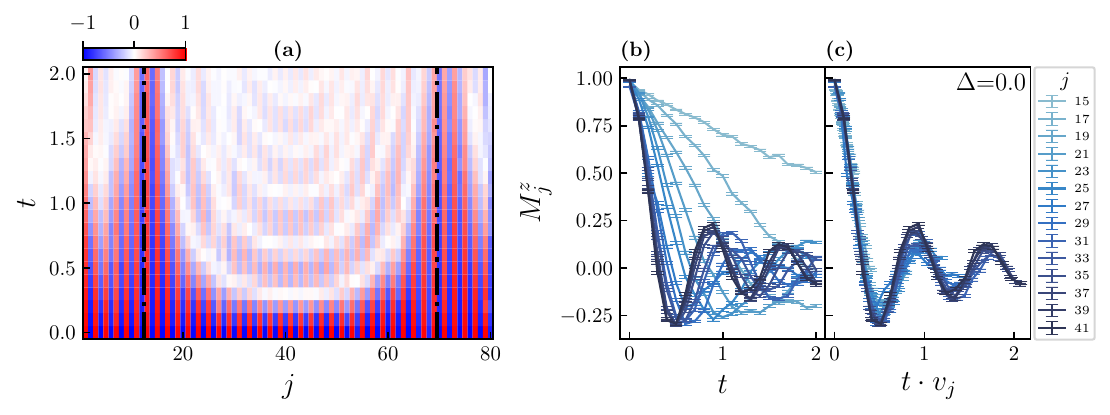}
    \caption{
    (a) data for the magnetization dynamics starting from a Néel state with $N=80$ qubits, evolving with the noninteracting inhomogeneous spin chain, $\Delta=0$. (b) Damped oscillations at different initial positions reveal a strong position dependence of the frequency. (c) Collapse of the magnetization curves after rescaling the time axis by the spatial deformation.
    }
    \label{fig:delta0magnetization}
\end{figure}

We now present data on the magnetization dynamics for the XX chain, as shown in \cref{fig:delta0magnetization}. Similarly to the interacting setting, we observe a clear spatial dependence of the periodicity that reflects the deformation profile. We note that the oscillations are more robust compared to the interacting gapless setting $\Delta\neq 0$ (shown in the main text in \cref{fig:magnetization_dynamics}(a-c)), as expected.

\section{Ballistic spreading in the XXZ chain (\texorpdfstring{$\Delta=1/2$}{Delta=1/2})}
\label{app:ballsiticspread}

In this section, we discuss the experimental procedure to extract the unequal-time spin-spin correlation function $G^{zz}_{ji}(t,0) = \langle \sigma^z_j(t)\sigma^z_i(0) \rangle$ along with the background subtraction method used to generate \cref{fig:ballisticcorrelator} in the main text.
Due to the form of the initial states considered in this work, \cref{eq:initial_states}, the unequal-time spin-spin correlation function is related to the local magnetization by
$G^{zz}_{ji}(t,0) = M^z_j(t) M^z_i(0)$ (assuming perfect initial state preparation).
Furthermore, one should observe $M^z_i(0)=\pm 1$, which simply contributes an overall sign. Hence, moving forward, we focus only on $M^z_j(t)$ as an effective proxy for $G^{zz}_{ji}(t,0)$.
Using $N=80$ qubits on \texttt{ibm\_fez}, we consider various initial states time evolving under the $\Delta=1/2$ XXZ Hamiltonian for $s=0,1,2,\dots,20$ Trotterized steps, with both uniform couplings and deformed couplings given by the profile in \cref{eq:deformationprofileXXZ}.
Upon reaching the discrete time $t_s = s\delta t$ with $\delta t = 0.1$, we measure the final state in the computational basis using $2^{14}$ shots.
From the shot data, we extract the local $z$-component of the magnetization.
For both uniform and deformed Hamiltonians, we consider the following initial states, $\ket{\Psi_0}$:
\begin{align}
    \label{eq:ballistic_initial_states}
    | \uparrow\dots\uparrow \rangle
    \quad,\quad
    | 20,30 \rangle
    = \sig^x_{20} \sig^x_{30}  | \uparrow\dots\uparrow \rangle
    \quad,\quad
    | 20 \rangle
    = \sig^x_{20} | \uparrow\dots\uparrow \rangle
    \quad,\quad
    | 30 \rangle
    = \sig^x_{30} | \uparrow\dots\uparrow \rangle
    .
\end{align}
\textit{Note}: In order for the measured observables to experience similar hardware noise across all initial states, these jobs were submitted in quick succession.
In the specific case of the deformed XXZ chain with the all-up initial state, $| \Psi_0 \rangle = | \uparrow\dots\uparrow \rangle$, the $s=14$ Trotter step job submission failed to run.
Due to the background subtraction method to be outlined in what follows, and without running all simulations again, we opted to approximate the $s=14$ correlators for this particular run using $G(t=14 \delta t) \approx \frac{1}{2}\left[ G(t=13 \delta t) + G(t=15 \delta t) \right]$ where $G(t)$ represents an arbitrary equal-time correlation function.

\begin{figure}[htbp]
    \centering
    \includegraphics[width = \textwidth]{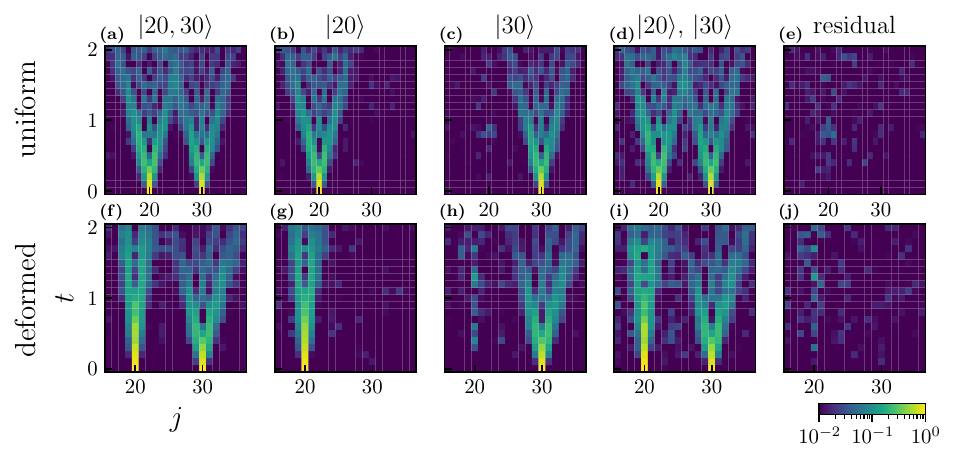}
    \caption{
    (a-e):
    Local magnetization data generated from quantum simulations of the uniform $\Delta=1/2$ XXZ chain performed on \texttt{ibm\_fez} with $N=80$ qubits.
    (f-j): Quantum simulations using the same parameters as the top row, except
    now with the deformation profile in \cref{eq:deformationprofileXXZ}.
    Columns 1, 2, and 3 correspond to the background-subtracted local magnetization, \cref{eq:ballistic_protocol_1},
    for initial states $\ket{\Psi_0} = | 20,30 \rangle , | 20 \rangle , | 30 \rangle$ respectively.
    Column 4 shows the results of adding together the local magnetization produced from initial states $\ket{\Psi_0} = | 20 \rangle$ and $\ket{\Psi_0} = | 30 \rangle$ (including the background subtraction), \cref{eq:ballistic_protocol_2}.
    Column 5 displays the absolute value of the difference between the data sets in column 1 and 4.
    Consistent log scale color maps are used throughout whose minimum value, $10^{-2}$, was chosen as it is on the same order of magnitude as the standard error in the local magnetization.
    }
    \label{fig:ballistic_magnetization_residuals}
\end{figure}

For the $\Delta = 1/2$ XXZ chain, we expect correlations to spread ballistically, as they are carried by effectively noninteracting gapless quasiparticles.
To probe this, we consider the behavior of the local magnetization generated from two distinct protocols.
In the first protocol, we measure the local magnetization in the case where the initial state has 2 spins pointing down at sites 20 and 30, $M^z_j(t;| 20,30 \rangle)$.
In the second protocol, we measure the local magnetization in the case where the initial state has a single down spin at site 20, $M^z_j(t;| 20 \rangle)$, and also perform the same experiments for an initial state that has a single down spin at site 30, $M^z_j(t;| 30 \rangle)$. We then add these correlators, $M^z_j(t;| 20 \rangle) + M^z_j(t;| 30 \rangle)$.

For information that spreads ballistically, we expect the same results from the two protocols.
However, because of the noise in the quantum hardware, one should expect that the additive nature of the second protocol will amplify systematic noise.
To mitigate this, we perform a background subtraction using the local magnetization measured in the case where the initial state has all the spins pointing up, $M^z_j(t;| \uparrow\dots\uparrow \rangle)$.
Because the all-up state $| \uparrow\dots\uparrow \rangle$ is an eigenstate of the XXZ Hamiltonian, in noise-free hardware the magnetization would always be equal to $1$.
With this in mind, in the first, second, and third columns of \cref{fig:ballistic_magnetization_residuals}, we show the background-subtracted local magnetization for the initial states of interest:
\begin{align}
    \label{eq:ballistic_protocol_1}
    \frac{1}{2} \left| M^z_j(t;| \Psi_0 \rangle) - M^z_j(t;| \uparrow\dots\uparrow \rangle) \right|
    ,
\end{align}
where $| \Psi_0 \rangle = | 20,30 \rangle, | 20 \rangle, | 30 \rangle$ and the range of the function is the unit interval $[0,1]$.
For both uniform and deformed chains, we have zoomed in on the combined causal regions of the excitations emitted at sites 20 and 30.
In the fourth column of \cref{fig:ballistic_magnetization_residuals}, we show the background-subtracted local magnetization when the $| 20 \rangle$ and $| 30 \rangle$ initial state experimental runs are added together:
\begin{align}
    \label{eq:ballistic_protocol_2}
    \frac{1}{2} \left |M^z_j(t ; | 20 \rangle) + M^z_j(t ; | 30 \rangle) - 2 M^z_j(t ; | \uparrow\dots\uparrow \rangle) \right|
    .
\end{align}
Finally, in the last column of \cref{fig:ballistic_magnetization_residuals}, we quantify the similarity of the two protocols (\cref{eq:ballistic_protocol_1} for $\ket{\Psi_0} = \ket{20,30}$ and \cref{eq:ballistic_protocol_2}) by displaying the absolute value of the difference between the two data sets.
In both uniform and deformed chains,
we see excellent agreement between the two approaches in the causal region, with the strongest deviations occurring near site 20.

\section{Supplementary data for the connected two-point correlation function}
\label{app:twopoint}
\begin{figure*}[!htbp]
    \centering
    \includegraphics[width = \textwidth]{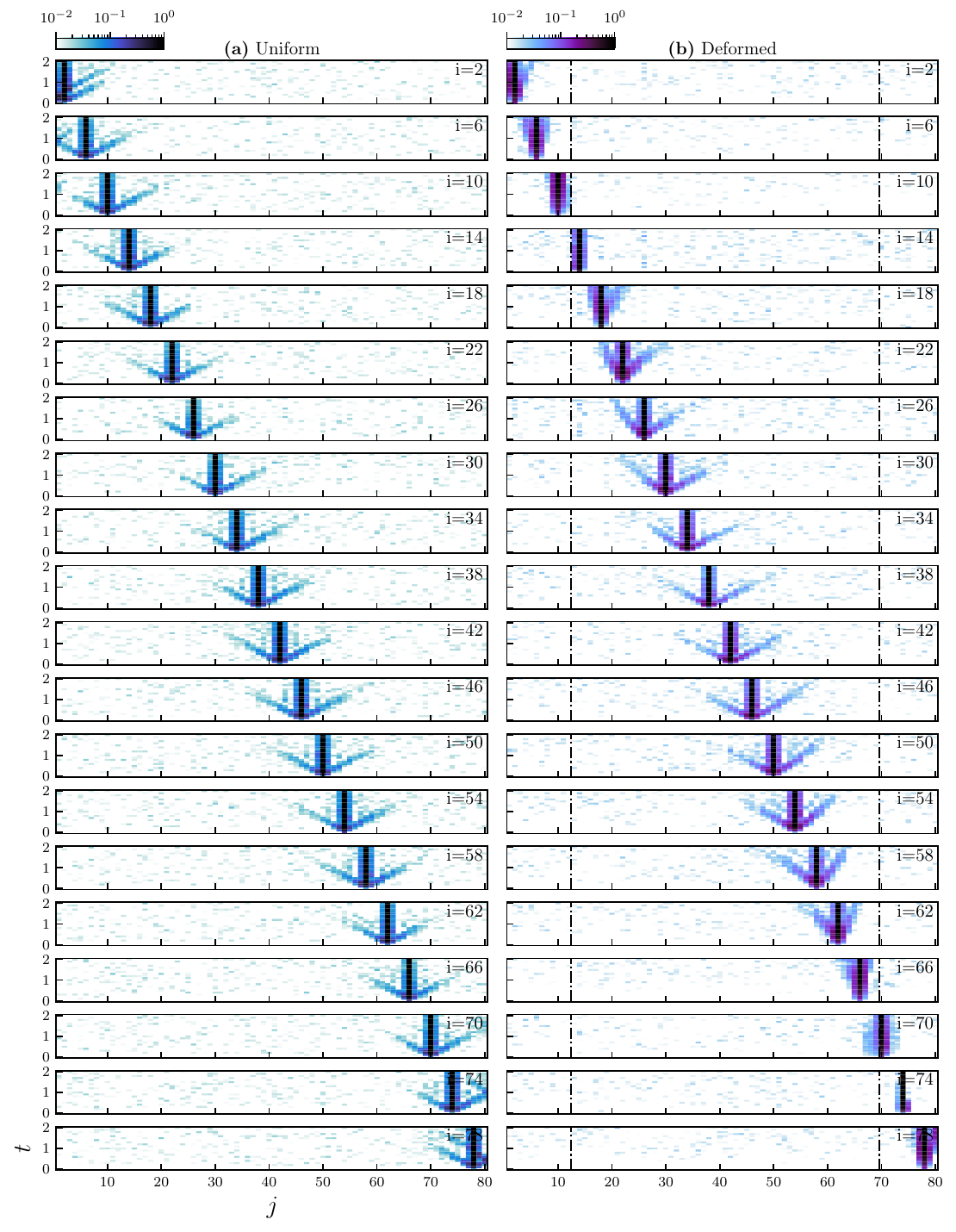}
    \caption{
    Digital quantum simulations of the $\Delta = 1/2$ XXZ chain using $N=80$ qubits initialized in a Néel state.
    (a) Uniform chain simulation run on \texttt{ibm\_fez} on 2025-05-30.
    (b) Deformed chain simulation run on \texttt{ibm\_marrakesh} on 2025-05-30.
    The deformation profile in (b) is given by \cref{eq:deformationprofileXXZ} and Rindler horizons located at $j_*+1$ and $N+1-j_*$ ($j_*=N/7$) are indicated with black dash-dot vertical lines.
    Here we show $|C^{zz}_{ij}(t)|$ data for every 4th site in the chain.
    }
    \label{fig:Neel_N_over_seven_Delta_0.5}
\end{figure*}

\begin{figure*}[!htbp]
    \centering
    \includegraphics[width = \textwidth]{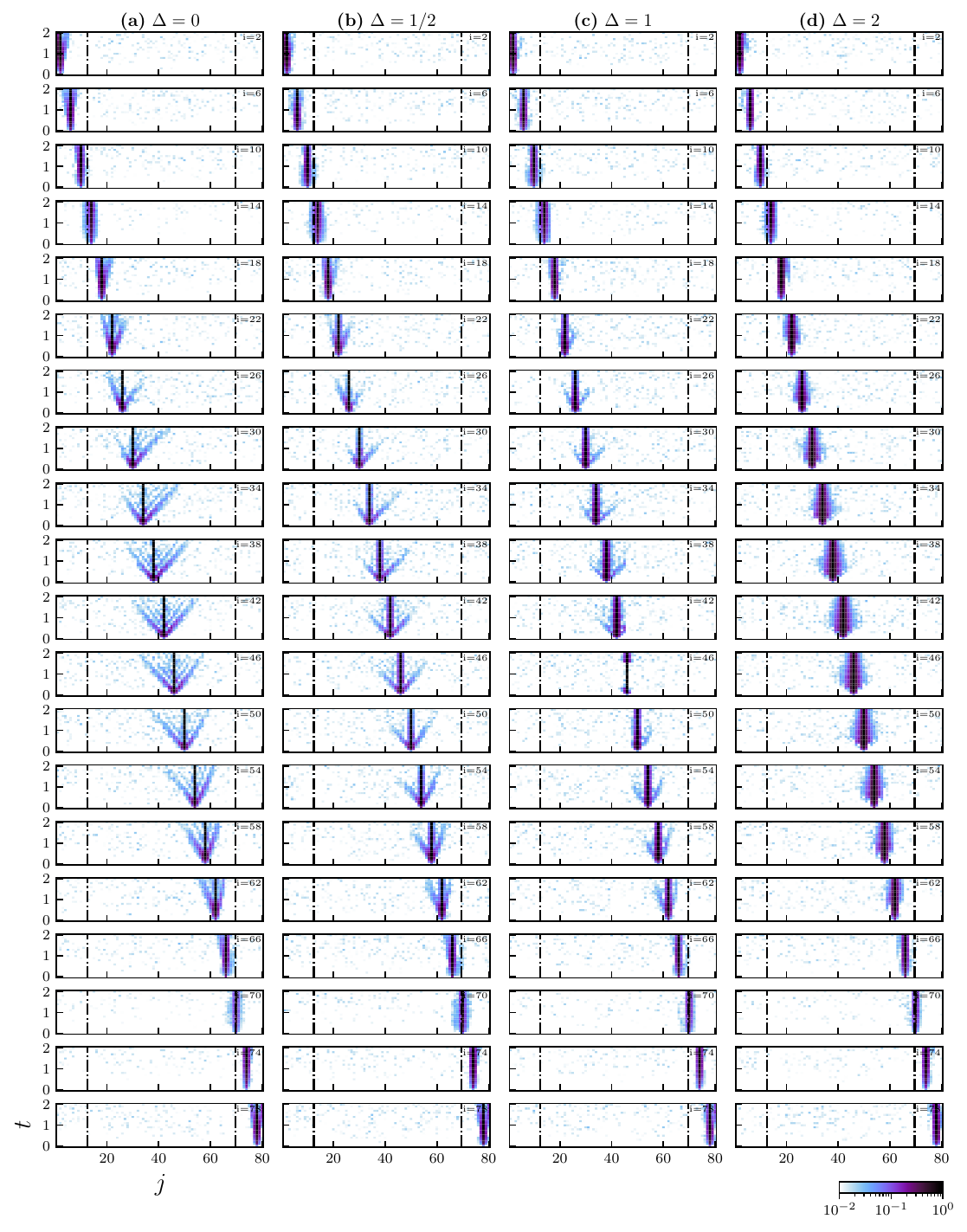}
    \caption{
    Digital quantum simulations of the inhomogeneous XXZ chain using $N=80$ qubits initialized in a Néel state and various anisotropies.
    All simulation were run on \texttt{ibm\_fez} on 2025-05-30.
    Columns (a)-to-(d) correspond to simulations using $\Delta = 0,1/2,1,2$ respectively.
    The deformation profile is given by \cref{eq:deformationprofileXXZ} and Rindler horizons located at $j_*+1$ and $N+1-j_*$ ($j_*=N/7$) are indicated with black dash-dot vertical lines.
    Here we show $|C^{zz}_{ij}(t)|$ data for every 4th site in the chain.
    }
    \label{fig:Neel_N_over_seven_varying_Delta}
\end{figure*}

On the following pages, we provide additional two-point correlation function data for the case that a $N=80$ qubit chain, initialized in a Néel state, is quenched according to the XXZ Hamiltonian, \cref{eq:deformedXXZ}.
In \cref{fig:Neel_N_over_seven_Delta_0.5}, expanding on the data shown in \cref{fig:curvedlight-cone_propagation} of the main text, we show the dynamics generated by both uniform and deformed Hamiltonians for the case that $\Delta=1/2$.
In \cref{fig:Neel_N_over_seven_varying_Delta}, expanding on the data shown in \cref{fig:different_Delta} of the main text, we show the dynamics generated by the deformed Hamiltonian for different values of the anisotropy.
For all inhomogeneous XXZ chains, the deformation profile is given by \cref{eq:deformationprofileXXZ} with $j_*=N/7$.

\newpage

\end{document}